%% file: prospects.tex
\documentstyle[epsfig,a4,12pt]{article}
\epsfverbosetrue

\def\nuebar{\bar{\nu_e}}
\def\dm2{\rm{\Delta m^2}}
\def\munu{\mu_{\nu}}
\def\nue{\nu_{e}}
\def\snu{\rm{\nu_{\odot}}}

\def\s2tw{\rm{ sin ^2 \theta _W }}
\def\am241{\rm{ ^{241} Am }}
\def\u238{\rm{ ^{238} U }}
\def\th232{\rm{ ^{232} Th }}
\def\k40{\rm{ ^{40} K }}

\begin{document}

\hfill AS-TEXONO/99-02\\
%%\hspace*{1cm} \hfill hep-ex/9910002\\
\hspace*{1cm} \hfill Revised: March 21, 2000.

\begin{center}
\large
\bf{
Prospects of Scintillating Crystal Detector in\\
Low-Energy Low-Background Experiments\\
}
\vspace*{0.5cm}
\normalsize
H.T. Wong~$^{\alpha ,}$~\footnote{Corresponding Author. 
E-mail: htwong@phys.sinica.edu.tw}, 
J. Li~$^{\beta}$,
C.Y. Chang~$^{\gamma}$, C.C. Chang~$^{\gamma}$,
C.P.~Chen~$^{\alpha}$,
W.P.~Lai~$^{\alpha}$,\\ 
H.B.~Li~$^{\delta}$, Y.~Liu~$^{\beta}$, 
J.G. Lu~$^{\beta}$, Z.P. Mao~$^{\beta}$,
S.C.~Wang~$^{\alpha}$\\
%%, S.Q.~Zhao~$^{\beta}$\\
\end{center}
\vspace*{0.05cm}
\begin{flushleft}
$^{\alpha}$ Institute of Physics, Academia Sinica, Taipei 11529, Taiwan.\\
$^{\beta}$ Institute of High Energy Physics, Beijing 100039, China.\\
$^{\gamma}$ Department of Physics, University of Maryland, 
Maryland 20742, U.S.A.\\
$^{\delta}$ Department of Physics, National Taiwan University, 
Taipei 10617, Taiwan.
\end{flushleft}
\vspace*{0.5cm}
\begin{center}
{\bf
Abstract
}
\end{center}

Scintillating crystal detector offers
potential advantages in low-energy (keV-MeV range)
low-background
experiments for particle physics and astrophysics.
The merits are discussed using CsI(Tl) crystal
as illustrations.
The various physics topics which can be pursued
with this detector technology are summarized.
A conceptual design for  a generic detector
is presented.

\vspace*{0.1cm}

\begin{flushleft}
{\bf PACS Codes:} 14.60.Pq; 95.35.+d; 29.40.Mc \\
{\bf Keywords:} Neutrinos; Dark Matter; Scintillation detector
\end{flushleft}

\begin{center}
\vfill {\it (published in Astroparticle Physics 14, 141 (2000).) }
\end{center}

\newpage

\section{Introduction}

Scintillating crystal detectors~\cite{crystal} have been
widely used as electromagnetic calorimeters in
high energy physics~\cite{emcalo}, as well as in medical and
security imaging
and in the oil-extraction industry.
They have  also been adopted for non-accelerator experiments,
notably NaI(Tl) detectors are already
used in Dark Matter searches~\cite{dmnai,dama}, producing some of the
most sensitive results.

Several characteristic properties
make crystal scintillator an
attractive detector option for low-energy (keV to MeV
range) low-background experiments.  
Subsequent sections of this article will
bring out the potential advantages of this
approach and some of the physics
topics well-suited to be investigated
by this detector technology.
The characteristic performance of the CsI(Tl) crystal
are used as illustrations.
A generic design to exploit these merits
in a realistic experiment
is discussed in Section~\ref{sect::design}.

\section{Motivations and Merits}
\label{sect::merit}

\subsection{Nuclear Physics}

The physics at the keV to MeV range can 
depend critically on the choice of isotopes
as the interaction targets.
The nuclear structure  determines the
interaction cross-sections, detection
threshold, as well as experimental signatures
like spatial or temporal correlations.

The deployment of target with large mass 
for low energy experiments
usually requires  that the target is an active
detector.
There are only a few
detector technologies which accommodates a wide range of possible
nuclei. The choice is even more limited when the
potentials and possibilities to scale-up to 
a massive (tons or more) detector have to be
taken into account.
The two most prominent candidate techniques
are loaded liquid scintillator and crystal
scintillator.

\subsection{Existing Experience and Potential Spin-Offs}

The large target-mass requirement 
is always a challenge
to low count rate experiments. 
From the big electro-magnetic calorimeters
in high energy physics, there are 
much experience in
producing and operating 50-ton-range crystal 
calorimeters.
The technology is proven to be stable and robust at
the harsh accelerator environment.
Indeed, the present
broad applications and affordable
price range for crystals like CsI(Tl) and BGO
are driven mostly by the demand and development
from high energy physics experiments.
Therefore, it is possible that construction
of a big scintillating crystal detector
for low-energy experiments will also lead
to the maturity of a new technology
with potential spin-offs in other areas.

\subsection{Intrinsic Properties}

Some of the properties of crystal scintillators
make them favorable candidates to be adopted
for studying low-energy low-background 
experiments {\it relative}
to the various other proposed detector schemes. 
While different crystals do have
different characteristic performance parameters,
the merits of this detector approach
are discussed below using CsI(Tl) as
an illustration. 

The characteristic properties
of CsI(Tl)~\cite{csichar}, together with a few other 
common crystals as well as liquid
and plastic scintillators, are summarized 
and compared in Table~\ref{scintab}.
The selection of CsI(Tl) is due to the
fact that it is a commonly used 
and relatively inexpensive crystal
scintillator produced in large quantities
and with many examples of successful
operation as 50-ton-range electro-magnetic
calorimeters, as in all the B-factory detectors 
currently under operation.
Unlike NaI(Tl), it is
only slightly hygroscopic and can operate
stably for a long time
without the need of a hermetic seal (based
on experiences from high energy physics experiments).
This minimizes the use of passive materials at the
target volume which,
as explained below, is crucial to allow
the merits of this detector technique 
for low-energy low-background experiments
to be fully exploited.

\subsubsection{Solid and Compact Detector}

Crystal scintillators usually have 
high density and are made up of high-Z
isotopes. Therefore, a massive (tens
of tons) detector can still be very
compact (scale of several m$^3$), such
that external shielding configurations
can be made more efficient and 
cost effective. The compact dimension
also favors applications where 
artificial neutrino sources are
used thereby allowing efficient
exposure of the target materials
to the source.

A solid detector can also prevent radioactive
radon gas from diffusing into the inner fiducial
volume from the external surfaces. This is a
major concern for target based  on gaseous 
or liquid detectors. Special procedures are still
necessary to minimize the radon contaminations
on crystal surfaces, as noted in Section~\ref{sect::focus}
and Section~\ref{sect::design}.

\subsubsection{Efficient  Active Veto} 

\begin{figure}
\centerline{
\epsfig{file=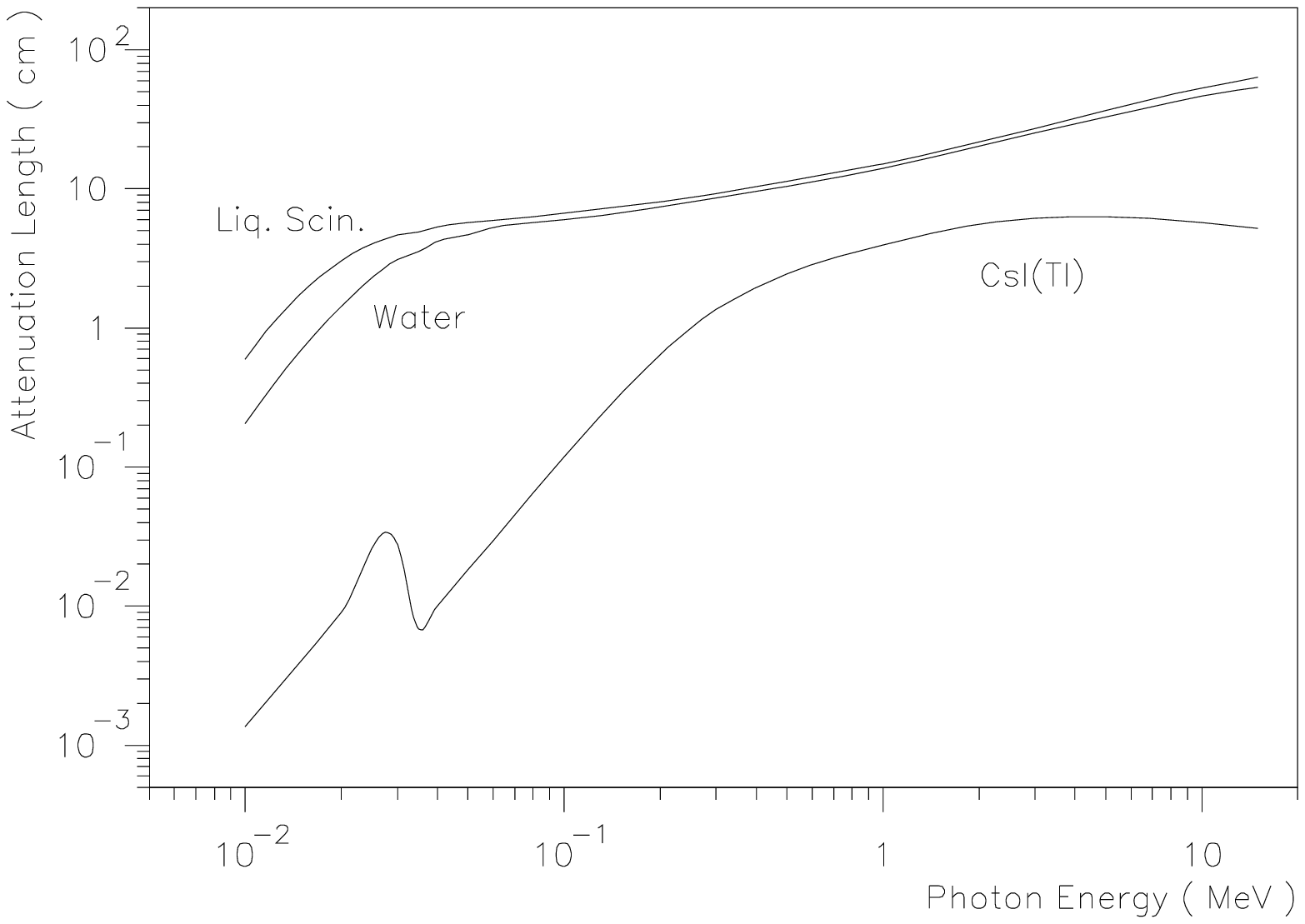,width=15cm}
}
\caption{
The attenuation length, as defined by the interactions
that lead to a loss of energy in the media,
for photons at different energies,
for CsI(Tl), water, and liquid scintillator.
}
\label{attgamma}
\end{figure}

The attenuation effects
of CsI(Tl) to $\gamma$-rays of
different energy, together with those of water and
liquid scintillator (a generic CH$_2$ compound with
density 0.9~gcm$^{-3}$)~\cite{gamal}, are depicted in 
Figure~\ref{attgamma}. 
In the region between 500~keV
to around 3~MeV, Compton scattering (which varies
as the atomic number Z) is the main process, 
and therefore the
attenuation effects of CsI(Tl) are only enhanced by
the density ratio, relative to H$_2$O and CH$_2$.
Above several MeV, pair production (varying as Z$^2$)
takes over making the high-Z CsI(Tl) more efficient.
This is the reason of choosing this crystal as
electromagnetic calorimeters. 
In the low energy region below 500~keV, photo-electric
effects (varying as Z$^5$) dominates overwhelmingly.
For instance, the attenuation lengths for a 100~keV
$\gamma$-ray are 0.12~cm and 6.7~cm, respectively,
for CsI(Tl) and CH$_2$. That is, 
1~cm of CsI(Tl) is equivalent to 8 attenuation lengths,
and 10~cm of CsI(Tl)
has the same attenuating power as 5.6~m of liquid
scintillator  at this low energy. Most crystal
scintillators, having high-Z isotopes,
share this merit.

\subsubsection{Focus only on Internal Background} 
\label{sect::focus}

Given the large attenuating effects
on low energy photons, 
crystal detectors can  provide a unique advantage
to the background suppression in low energy
experiments - that {\it external $\gamma$-background
are highly suppressed such that practically
all $\gamma$-background originates
internally IF (1) a three-dimensional fiducial
volume can be defined, and (2) a housing-free
design with minimal passive materials can be realized.}

For non-hygroscopic crystals like
CsI(Tl) or BGO where a hermetic seal system
is not
needed for their operation, 
{\it ``internal''} would include
only two materials: the crystal itself and the
surface wrapping or coating materials. 
Teflon wrapping sheets are most commonly used,
while there is an interesting new development
with sol-gel coating which can be as thin
as a few microns~\cite{solgel}.
Teflon is known
to have very high radio-purity (typically
better than the ppb level for the
$^{238}$U and $^{232}$Th series)~\cite{hpge}. 

The suppression of radon contamination to the 
inner fiducial volume
requires special but standard procedures.
Crystal surfaces as well as the teflon wrapping 
sheets
should be cleaned before wrapping, preferably in
a nitrogen-filled glove-box. The detector modules
should be covered and protected by an additional
surface (like aluminium foils), which will be removed
only at the time of installation. The whole detector should
be installed and operate
in an air-tight box filled with clean nitrogen.

As a result, 
the experimental challenges and hurdles become
focussed on to two distinct aspects:
\begin{description}
\item[{\bf I. Background-wise:}]
the control and understanding
of the internal purity of the crystal target itself; and
\item[{\bf II. Detector-wise:}]
the realization of a detector design giving good position
resolutions and with as low a threshold as possible.
\end{description}
Accordingly, the difficulties for
external gamma-background control 
can be alleviated
at the expense of additional
detector requirements. 

The internal background can be due to
contaminations of naturally occurring isotopes
($^{238}$U and $^{232}$Th series, $^{40}$K),
long-lived fission products and cosmic ray-induced
unstable nuclei.
The background due to external $\gamma$'s, like those from 
the readout device, 
electronics components, 
construction materials, or
radon contamination on the outer surfaces, 
can thus be attenuated and vetoed by
the outer active volume.
Background can also originate externally from 
cosmic-ray induced neutrons which have
little attenuation with high-Z nuclei.
Their effects, however, can be minimized by
a cosmic veto and by operating the experiment
underground. 

Hygroscopic crystals like NaI(Tl) are housed in 
containers as hermetic seal.
The containers, usually made of oxygen-free
copper for low background application, can
be made to have high radio-purity. However, 
it is still an inactive material with high
cross-sections for photons. 
Consequently, it is possible that
high energy photons (which have
less attenuation in the crystal) can
penetrate into the fiducial volume,
undergo Compton scatterings at the
passive container, and deposit only low energy at
the crystal detector itself.
Therefore, the adoption
of non-hygroscopic crystals (that is, a housing-free set-up)
is essential to exploit the full power of this
merit - that there is big suppression
for external photons at low energies ($<$500~keV)
to get into fiducial volume or for 
those at high energies ($>$ MeV) to get into 
fiducial volume but deposit only 100~keV
of visible signals.

The background count rate
will be stable and not affected
by external parameters
if the dominant contributions are from
internal contaminations.
Therefore, this detector technique can 
provide additional desirable feature
in  applications requiring delicate comparison
and subtraction of data taken at different
periods (such as reaction ON/OFF, annual modulation,
Day/Night effects). The light yield for scintillating
crystals is usually temperature-dependent, and therefore
a good calibration scheme and temperature-control
of the detector region is
crucial to realize these subtraction procedures.

It should be stressed that an experimental design
which provides the definition of a three-dimensional
fiducial volume is essential to allow 
this large suppression of the external gamma background.
The extent to which this can be achieved in a realistic
detector set-up will depend on the specific crystal
properties (particularly the light yield), and the energy
range of interest. This will be discussed further 
in Sections~\ref{sect::design} and \ref{sect::bkg}.

\subsubsection{Good Energy Resolution and Modularity}

Light yield of typical crystal scintillators are
comparable to those of liquid and plastic scintillators.
However, the modular size
are smaller while the refractive index higher, 
leading to more efficient
light transmission and collection. The
high gamma attenuation also allows full
$\gamma$-energy deposition. Consequently,
crystal scintillators have typically better
energy resolution and lower detection
threshold, both of which are necessary for low-energy
measurements.
The high $\gamma$-rays
capture efficiency, together with the
good resolution to measure them as energy-peaks,
can  provide important diagnostic tools for understanding
the physical processes and 
background of the system.
For instance, by measuring the $\gamma$-peaks
due to $^{40}$K, $^{60}$Co and $^{137}$Cs, 
their associated $\beta$-background can be
accurately accounted for and subtracted off.

The good modularity also enhances
background suppression, since 
the interesting signals for most
applications are single-site
events. Most background from
internal radioactivity come as
$\beta$+$\gamma$'s in coincidence 
(like decays of $^{214}$Bi and $^{208}$Tl
from the $^{238}$U and $^{232}$Th series,
respectively) and hence will produce
multiple hits with high probability.
Similarly neutron capture events by the target isotopes
manifest as (n,$\gamma$) interactions,
giving rise to a $\gamma$-burst of multiple hits
with known total energy. 
The neutron capture rate can therefore be measured,
so that the background due to subsequent decays
of the unstable daughter nuclei can be subtracted off.

\subsubsection{Possibility of Pulse Shape Discrimination}

\begin{figure}
\centerline{
\epsfig{file=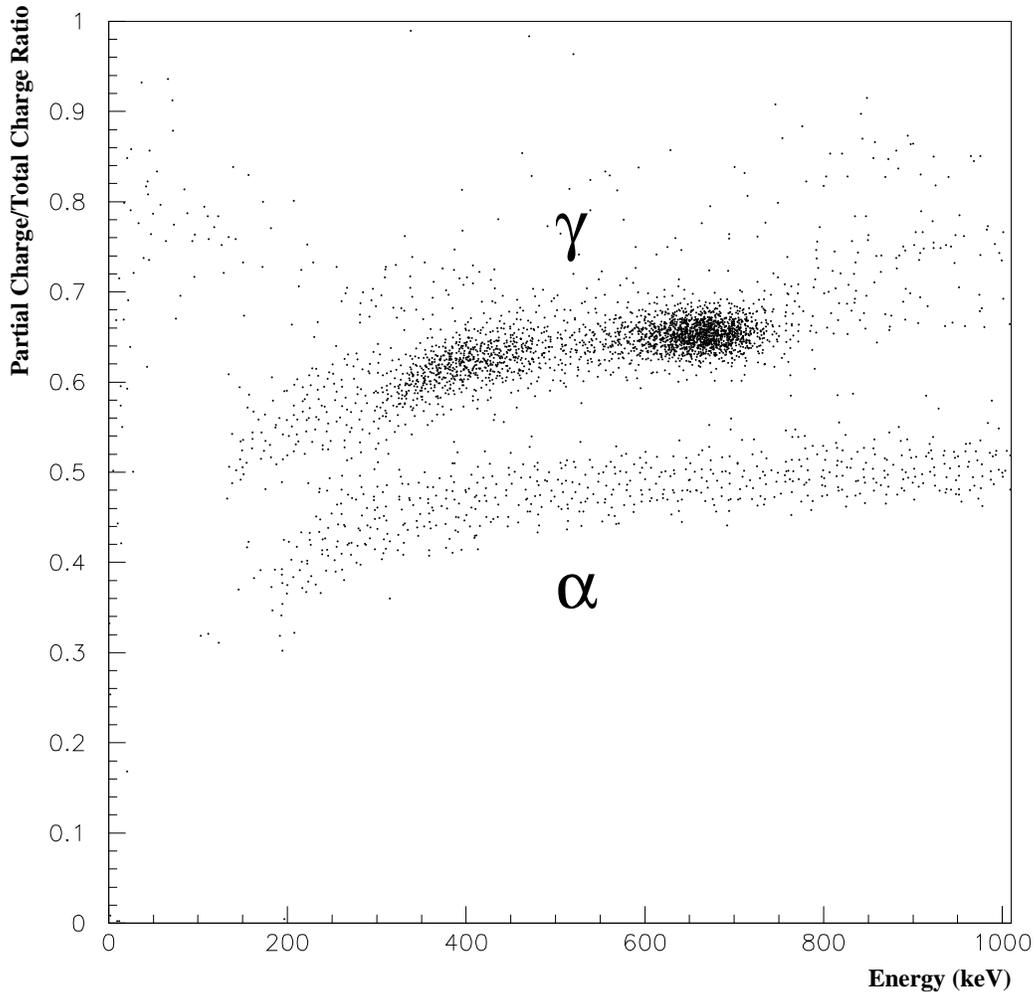,width=15cm}
}
\caption{
The partial charge/total charge ratio in a CsI(Tl) 
crystal as a function of energy, 
showing excellent pulse shape discrimination
capabilities to differentiate events due to
$\alpha$'s and $\gamma$'s. The $\gamma$-events are
due to a $^{137}$Cs source, showing peaks at the
full-energy and Compton edge regions.
The $\alpha$-events are from the low-energy tails of 
an $^{241}$Am source placed on the surface of the
crystal.
}
\label{psdspect}
\end{figure}

Crystals like CsI(Tl) and NaI(Tl) have superb
pulse shape discrimination (PSD) properties~\cite{csipsd}
to differentiate $\gamma$/e events from those
due to heavily ionizing particles like $\alpha$'s,
which have faster fall time. Figure~\ref{psdspect}
depicts typical PSD between $\alpha$/$\gamma$ in
CsI(Tl) with the ``Partial Charge Vs Total Charge''
method~\cite{psdmethod}, demonstrating excellent separation. 
The PSD capabilities provide powerful handle to
tag and study those background channels involving
alpha-emission, such as those from the
$^{238}$U and $^{232}$Th decay chains.

\subsubsection{High Sensitivity to U/Th Cascades}
\label{sect::alpha}

Unlike in liquid scintillators, $\alpha$'s are
only slightly quenched in their light output in
crystals like CsI(Tl) and NaI(Tl). The
exact quenching ratio depends on the Tl concentration
and the measurement parameters like shaping time: 
for full integration of the signals, 
the quenching is about 50\%~\cite{csichar}. 
Therefore, some of 
the $\alpha$'s emitted from the uranium and thorium
series are above 3~MeV. This is beyond the
end-point for natural radioactivity (2.61~MeV) and
hence the
peak signatures are easy to detect
among the flat background.
In comparison, the electron-equivalence light
yield for several MeV $\alpha$'s in liquid scintillators
is typically less than 10\% of their 
kinetic energy, making the signals well
below the natural end-point and therefore more
difficult to detect.

\begin{figure}
\centerline{
\epsfig{file=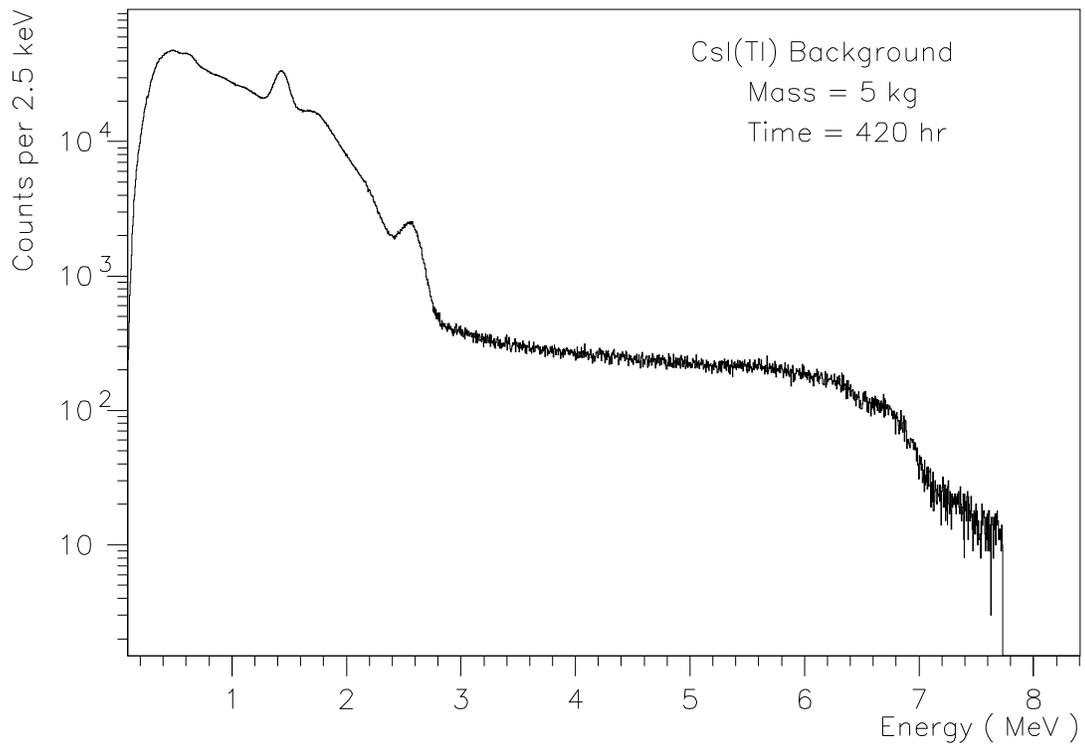,width=15cm}
}
\caption{
Measured background spectrum of a 5 kg CsI(Tl) under lead
shieldings and cosmic-ray veto. The absence of peaks 
above 3~MeV provides sensitive limits to contaminations
of $^{238}$U and $^{232}$Th decay chains in equilibrium.
}
\label{bkgspect}
\end{figure}

A crystal contaminated by uranium or thorium would
therefore give rise to multiple peaks above 3~MeV,
as reported in Ref.~\cite{csibkg} in the case
for CsI(Tl). Shown in Figure~\ref{bkgspect}
is the background spectrum from a 5-kg CsI(Tl) crystal
put in 5~cm of lead shielding with cosmic veto in
a typical sea-level laboratory. The absence of
multiple peaks above 3~MeV 
suggest a $^{238}$U and $^{232}$Th
concentration of less than
the $10^{-12}$~g/g level, assuming the
decay chains are in equilibrium. 
All the peaks and structures in the spectrum 
can be explained by ambient radioactivity or
by (n,$\gamma$) interactions at the crystal
and shielding materials.
This simple yet
effective measurement for crystal scintillator
can be compared to the complicated
schemes requiring elaborate underground facilities
for liquid scintillator~\cite{ctf}. A typical
level achieve-able by the photon-counting
method with a low-background 
germanium is only $10^{-9}$~g/g~\cite{hpge}.

The sensitivities can be pushed further
by doing the measurement underground (the flat
background above 3~MeV are due to 
cosmic-ray induced neutrons which undergo
(n,$\gamma$) when captured by the crystal
or the shielding materials), and 
by exploiting the PSD characteristics
of the crystal. In addition, by careful studies of the
timing and energy correlations among the $\alpha$'s, one
can obtain precise information on the
radioactive contaminants in the cases where
the $^{238}$U and $^{232}$Th decay series
are not in equilibrium, so that the associated
$\beta$/$\gamma$ background can be accounted for accurately.
For instance, some Dark Matter experiments with NaI(Tl)~\cite{dmnai}
reported trace contaminations (range of $10^{-18} - 10^{-19}$~g/g) 
of $^{210}$Pb in the detector, based on peaks from $\gamma$'s of 46.5~keV
and from $\alpha$'s of 5.4~MeV. Accordingly,
$\beta$-decays from $^{210}$Bi can be subtracted off
from the signal.

\section{Potential Applications}
\label{sect::phys}

Several areas of low energy particle physics where the
crystal scintillator technique may be applicable are
surveyed in this section. 

\subsection{Neutrino-Electron Scattering at Low Energy}

Scatterings of the
$\rm{( \nu_e ~ e )}$ and
$\rm{( \nuebar ~ e )}$ 
give information on the electro-weak parameters
($\rm{ g_V , ~ g_A , ~ and ~ sin ^2 \theta_W }$),
and are sensitive to
small neutrino magnetic moments ($\munu$)~\cite{vogelengel}.
They are two of the most
realistic systems
where the interference
effects between Z and W exchanges
can be studied~\cite{kayser}.

The goal of 
future experiments will be to push the
detection threshold as low as possible
to enhance the sensitivities in the 
magnetic moment search. 
Using reactor neutrinos as source, an experiment 
based on a gaseous time projection chamber
with CF$_4$~\cite{munu} is now operational.
Another experiment using CsI(Tl) is being built~\cite{texono},
with the goal of achieving a threshold of 100~keV.
Project with NaI(Tl) detector at an underground
site and using an artificial neutrino source 
has also been discussed~\cite{nuenai}.

\subsection{Neutral Current Excitation on Nuclei}

Neutral current excitation (NCEX) on nuclei by neutrinos has
been observed only in the case of $^{12}$C~\cite{karmennuex}
with 30~MeV neutrinos. Excitations with lower energies
using reactor neutrinos have been studied theoretically~\cite{nuex}
but not observed. 

Crystal scintillators, having good $\gamma$ resolution
and capture efficiency, are suitable to study these processes where
the experimental signatures are peaks in the energy spectra
with characteristic energies. Realistic experiments can be
based on using the crystal isotopes as active targets,
like $^{133}$Cs and $^{127}$I in CsI(Tl) or
$^{6}$Li, $^{7}$Li and $^{127}$I  in LiI(Eu). The $^{7}$Li case,
with a $\gamma$-energy of 480 keV,  has  particularly
large cross-sections. Alternatively, a compact passive
boron-rich target like B$_4$C can be inserted into an array of
CsI(Tl) detector modules~\cite{texono}. There are
theoretical work~\cite{nuexaxial}
suggesting that
the NCEX cross-sections on $^{10}$B and $^{11}$B 
are sensitive to the
axial isoscalar component of NC interactions
and the strange quark content of the nucleon.

\subsection{Dark Matter searches}

Direct searches of 
Weakly Interacting Massive Particles (WIMP)~\cite{dmreview}
are based on looking for the low-energy (few keV) nuclear recoil
signatures when they interact with the nuclei.
Crystal scintillators may offer an appropriate detector technique
for these studies from 
their PSD capabilities, as well as being
a matured technology where a large target mass 
is possible. The cross-sections depend on specific
isotopes~\cite{wimpcs}, based on their nuclear matrix elements and spin
states.

The NaI(Tl) crystal detectors~\cite{dmnai}
are already used in WIMP searches
Up to the scale of 100~kg target mass has been deployed~\cite{dama},
producing some of the most sensitive results. 
Other projects on  CaF$_2$(Eu)~\cite{caf2}
and CsI(Tl)~\cite{csiwimp} 
are also pursued.
In addition, 
searches have been performed~\cite{wimpncexpt} with
the WIMP-nuclei inelastic scattering~\cite{wimpncexth}
giving rise to NCEX. 

For crystal detectors
where a three-dimensional fiducial volume with
minimal passive materials can be defined,
there is no background due to external $\gamma$'s at this
low energy.
Internal $\beta$ background is suppressed
by the spectral distribution 
at this very low energy range.
For instance, less than 3$\times 10^{-4}$ of the 
$\beta$-decays in $^{40}$K (end-point 1.3 MeV), 
give rise to events below 10~keV.
However, achieving a three-dimensional fiducial volume
definition
will be more difficult at these low energies, as
elaborated in Section~\ref{sect::bkg}.

\subsection{Low Energy Solar Neutrinos}

The goal of future solar neutrino experiments~\cite{snureview}
will be to measure the low energy (pp and $^7$Be)
solar neutrino spectrum. Charged- (CC) and Neutral-current (NC)
on nuclei are attractive detection channels besides
neutrino-electron scattering. 
The CC mode can provide a direct measurement of the
$\nue$-spectrum from the Sun without
the convolutions necessary for the $\nu$-e 
channels, while the NC mode can provide
a solar model independent cross-check.
Crystal scintillators are possible means to realize 
detectors based on the CC and NC interactions.

Previously, crystals with indium 
has been investigated~\cite{snuin}
for a $\snu$-detector with $^{115}$In as
target which can provide a distinct
temporally and spatially correlated
triple coincidence signature.
Recently, the crystals LiI(Eu)~\cite{lii}
and GSO ($\rm{Gd_2 Si O_5 (Ce)}$)~\cite{lensgso}
are being considered. The attractive features
are that LiI(Eu) have large $\nue$N-CC
cross-sections for both $^{7}$Li and $^{127}$I,
and $\nue$N-NC (E$_{\gamma}$=480~keV) for $^{7}$Li,
while $^{160}$Gd in GSO can provide another time-delay
signature
for background suppression and for
tagging the flavor-specific $\nue$N-CC reactions.
The primary experimental challenge is the 
requirements of extremely
low background level due to the small
signal rate.

\subsection{Double Beta Decay}

The energy range of interest for the search of
neutrino-less double beta decay~\cite{bbreview}
is mostly above 1~MeV, and
hence some of the merits for crystal scintillators 
discussed in Section~\ref{sect::merit}
relative to the other techniques are no longer applicable.
We mention for completeness that there are efforts
on $^{115}$Cd with CdWO$_4$~\cite{bbcdwo4} 
and on $^{160}$Gd with GSO~\cite{bbgso} crystals.

\section{Generic Detector Design}
\label{sect::design}

To fully exploit the advantageous features discussed
in Section~\ref{sect::merit}, the 
design of a scintillating crystal detector
for low-energy low-background experiments
should enable the definition of a fiducial
volume with a surrounding active 4$\pi$-veto.

\begin{figure}
\centerline{
\epsfig{file=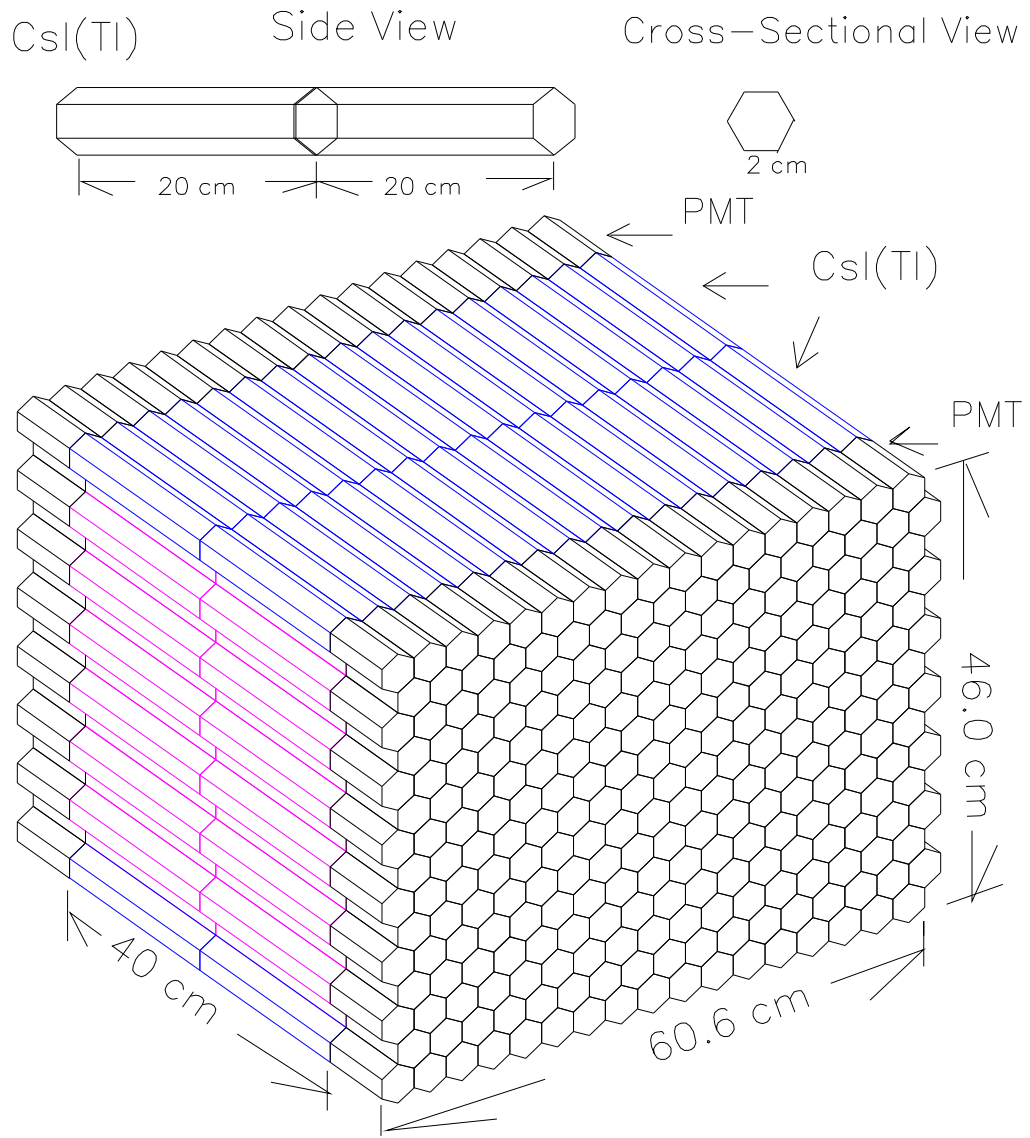,width=15cm}
}
\caption{
The schematic design of a generic
scintillating crystal detector.
}
\label{detdesign}
\end{figure}

Displayed in Figure~\ref{detdesign}
is a generic conceptual design where such a 
detector can be realized.
The detector design is 
based on an experiment being constructed~\cite{texono}
which, in its
first phase, will study low energy neutrino-electron
scattering from reactor neutrinos using CsI(Tl) 
as target. The listed dimensions are for 
this particular experiment.  The dimensions for
other  applications will naturally depend on 
the optimization based on the specific 
detector performance and requirements.

As shown in Figure~\ref{detdesign}, one CsI(Tl) crystal unit
consists of a hexagonal-shaped cross-section with 2~cm
side and a length 20~cm, giving a mass of 0.94~kg. 
Two such units are glued optically
at one end to form a module where the light output
from both ends are read out by photo-detectors.
Photo-multipliers (PMTs)
will be used for the experiment though
solid-state photo-detectors like photo-diodes or 
avalanche photo-diodes are also possibilities for other
applications. The modular design enables the detector
to be constructed in stages. 
Figure~\ref{detdesign} shows a
design with a 17$\times$15 matrix giving a total mass
of 480~kg.

The cleaning and wrapping procedures to minimize
radon contamination to the crystal surfaces
noted in Section~\ref{sect::focus}
will be adopted. The detector will operate inside an air-tight
box filled with dry nitrogen. The box itself will in turn 
be inside a nitrogen air-bag. 
The compact dimensions of the inner
target detector allow a more elaborate and cost-effective
shielding design. 
External to the air-bag are
the typical shielding configurations: from outside inwards
plastic scintillators for cosmic-ray veto, 15~cm of lead,
5~cm of steel, 25~cm of boron-loaded polyethylene, and
5~cm of copper. 
Potassium-free PMT glass window
as well as other high radio-purity materials will be
used near the target region.

\begin{figure}
\centerline{
\epsfig{file=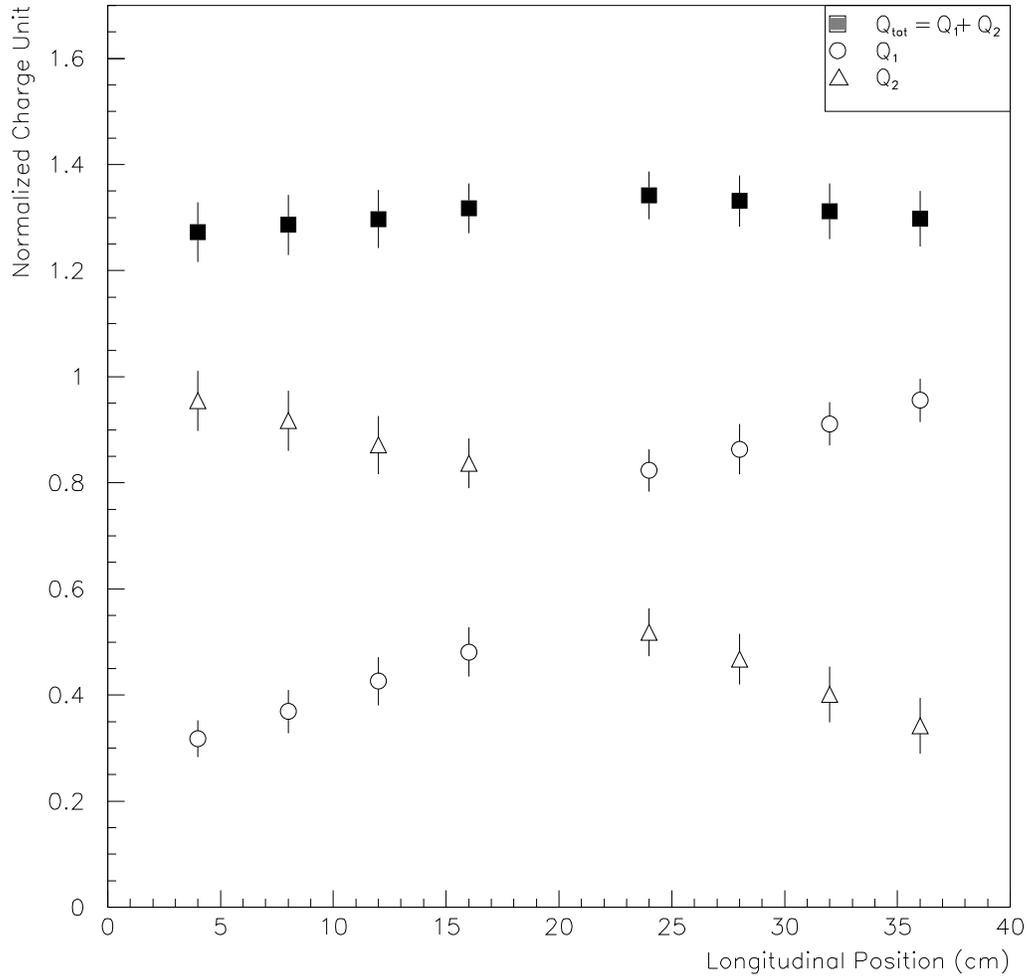,width=15cm}
}
\caption{
The measured variations of 
Q$_1$, Q$_2$ and $\rm{Q_{tot}= Q_1 + Q_2}$ along
the longitudinal position of the crystal module.
The charge unit is normalized to unity for both 
Q$_1$ and Q$_2$ at their respective ends.
The error bars denote the width  of the
photo-peaks due to a $^{137}$Cs source.
}
\label{qvsz}
\end{figure}

The energy deposited can be derived from the sum of
the two PMT output $\rm{ ( Q_{tot} = Q_1 + Q_2 ) }$ after
their gains are normalized, while the
longitudinal position can be obtained from their
difference in the form of 
$\rm { R = (  Q_1 - Q_2 ) / (  Q_1 + Q_2 ) }$.
The variation of $\rm{Q_1}$, $\rm{Q_2}$
and $\rm{Q_{tot}}$  along the crystal 
length are displayed in Figure~\ref{qvsz}.
The error bars denote the width of the photo-peaks
of a $^{137}$Cs source. 
The discontinuity in the middle is due to the optical
mismatch between the interface glue (n=1.5) and 
the crystal (n=1.8). It can be seen that $\rm{Q_{tot}}$ 
is independent of the position, and the resolution
at 660~keV is about 10\%. 
The detection threshold (where signals
are measured at both PMTs) is $<$20~keV.

\begin{figure}
\centerline{
\epsfig{file=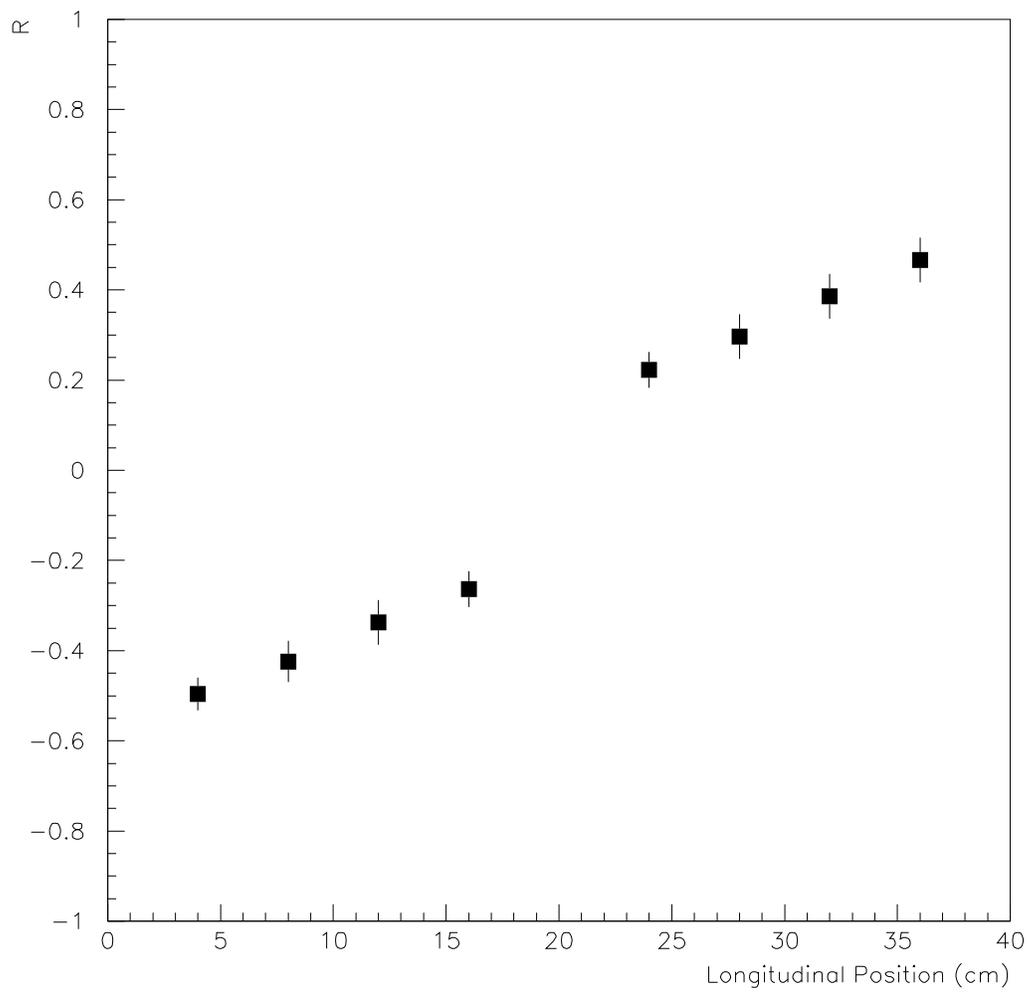,width=15cm}
}
\caption{
The variation of 
$\rm { R = (  Q_1 - Q_2 ) / (  Q_1 + Q_2 ) }$
along the longitudinal position of the crystal module,
showing the capability to provide a
position measurement.
}
\label{rvsz}
\end{figure}

\begin{figure}
\centerline{
\epsfig{file=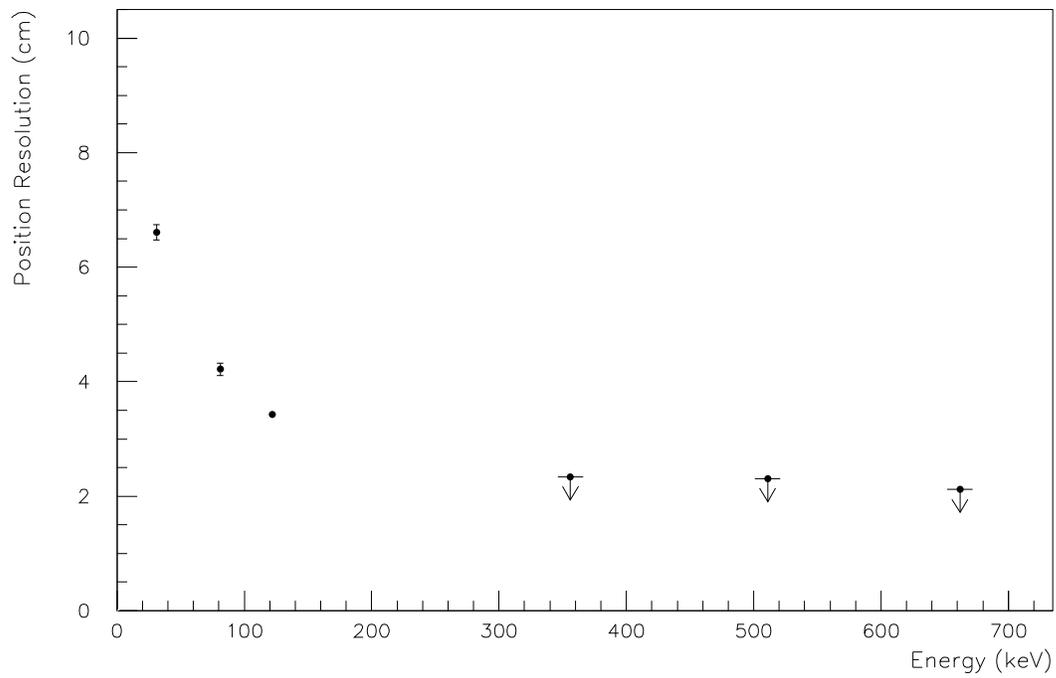,width=16cm}
}
\caption{
The variation of longitudinal position resolution
as a function of gamma-energy,  based on
measurements with $\gamma$-sources at different energies.
Limits are set for energy above 350~keV due to finite
collimator size and the non-locationization of the
multiple Compton scattering events.
}
\label{dzvse}
\end{figure}

The variation
of R along the crystal length
is depicted in Figure~\ref{rvsz}. 
The ratio of the RMS errors in R relative to the slope 
gives the longitudinal position resolution.
Its variation as a function of 
energy, obtained from measurements with $\gamma$-sources
of different energies,
is displayed in Figure~\ref{dzvse}, showing
a resolution of $<$2~cm and 4~cm at 660~keV and 100~keV, respectively.
Only upper limit 2~cm on the resolution 
can be concluded above 350~keV due to
(a) finite collimator size for the
calibration sources, and (b) the event-sites of
$\gamma$-interactions (mostly multiple Compton scattering)
being less localized at higher energies.
It can be seen that a three-dimension fiducial volume 
can be defined above 50~keV,
where the definitions can be optimized for different
energy ranges. For instance, a 10~cm active veto length 
will give a suppression factor of $5 \times 10^{-3}$ to
external photons of 100~keV.
The fiducial volume only consists of the crystal
itself and the teflon wrapping sheets, typically in a mass
ratio of 1000:1.

Individual modules will be calibrated, both in light yield
and the light transmission profile, before installation.
On site, the stability of the crystals and PMTs
can be monitored by radioactive sources
illuminating the two end surfaces, as well as by cosmic-ray events.
Stability of the electronics can be monitored with a precision
pulse generation. LEDs placed at the end surface near
the PMTs can be used to monitor stability
of PMTs' response as well as the light transmission through
the crystal.

The various potential experiments based on
scintillating crystal detectors
can essentially adopt a similar design.
Much flexibility is available for optimization.
Different modules can be made of different 
crystals. 
Different crystals can be glued together
to form ``phoswich'' detectors,
in which cases the event location 
among the various 
crystals can be deduced from the different 
pulse shape.  Passive target, as well as
a different detector technology,  can be inserted
to replace a crystal module. 

\section{Background Discussions}
\label{sect::bkg}

Background understanding is crucial in all
low-background experiments. It is beyond the scope
of this article to present a full
discussion of the background and 
sensitivities for 
all the possible candidate
crystals in the various potential applications 
listed in Section~\ref{sect::phys}. 
In this Section, we consider
the key ingredients in the background issues 
relating to the CsI(Tl) experiment~\cite{texono},
followed by discussions
of possible extensions to the lower-energy
or smaller signal-rate experiments.

The experiment will operate at a shallow depth (about 
30~meter-water-equivalent) near a reactor core, with
the goal of achieving an 100~keV physics threshold 
corresponding to a $\nuebar$-electron  signal rate
of O(1)~per~kg of CsI(Tl) per~day 
[$\equiv$  ``pkd'']\footnote{For simplicity,
we denote {\it ``events per kg of CsI(Tl) per day''} by
{\bf pkd}
in this section.} 
As noted from discussions in Section~\ref{sect::focus}
and detector 
performance parameters achieved in Section~\ref{sect::design},
while care and the standard
procedures should be adopted for suppressing the
ambient radioactivity background as well as
those from the equipment and surrounding materials,
the dominant background channel is expected to be
that of {\it internal} background
from the CsI(Tl) itself. 
Based on
prototype measurements as well
as detector and shielding simulations, 
the various
contributions are summarized below.

\begin{enumerate}
\item {\bf Internal Intrinsic Radioactivity:}\\
Figure~\ref{bkgspect} and discussions
in Section~\ref{sect::alpha}
demonstrate that a $^{238}$U and $^{232}$Th
concentration of less than
the $10^{-12}$~g/g level [$\sim$1~pkd],
assuming the decay chains are in equilibrium.
In addition,
direct counting method with a high-purity
germanium detector
shows the $^{40}$K and $^{137}$Cs
contaminations of less than the $10^{-10}$~g/g~[$\sim$1700~pkd]
and $4 \times 10^{-18}$~g/g~[$\sim$1200~pkd] levels, respectively.
Mass spectrometry method sets limits of $^{87}$Rb to
less than $8 \times 10^{-9}$~g/g~[$\sim$210~pkd].
\item {\bf Neutron Capture}\\
The important channel comes from (n,$\gamma$) on $^{127}$I
producing $^{128}$I~($\rm{ \tau_{\frac{1}{2}} = 25~min~;~ Q = 2.14~MeV }$). 
Ambient neutrons or those produced at the
the lead shieldings have little probability
of being captured by the CsI crystal target,
being attenuated efficiently by the boron-loaded
polyethylene. Neutron capture by the target are mostly due to
cosmic-induced neutrons originated from the target itself,
such that the $^{128}$I production rate is about
1.8~pkd.\\
The other neutron-activated isotope,
$^{134}$Cs ($\rm{\tau_{\frac{1}{2}} = 2.05~yr ~;~ Q = 2.06~MeV }$),
decays with 70\% branching ratio by
beta-decay (end point 658~keV),
plus the emission of two $\gamma$'s (605~keV and 796~keV),
and therefore will not give rise to
a single hit at the low-energy region.
The probability of producing single-hit at the 1.5-2~MeV region is
suppressed by a factor of $<$0.05.

\item {\bf Muon Capture}\\
Cosmic-muons can be stopped by the target nuclei and
subsequently captured~\cite{refmucapture}. 
The process will give
rise to $^{133}$Xe and $^{127}$Te ($<$0.05 probability),
both of which can lead to low-energy single-site background
events. The expected rate is less than 1.5~pkd.
The other daughter isotopes are stable.

\item {\bf Muon-Induced Nuclear Dissociation}\\
Cosmic-muons can disintegrate the target nuclei
via the ($\gamma$,n) interactions or by
spallations~\cite{refmudis}, at
an estimated rate of $\sim$10~pkd and $\sim$1~pkd, respectively.
Among the various decay configurations of
the final states nuclei of the ($\gamma$,n) processes,
$^{132}$Cs and $^{126}$I, only about
20\% (or $\sim$2~pkd) of the cases will give rise
to low-energy single-hit background.
\end{enumerate}

Therefore, the present studies place limits on internal
radio-purity to the range of less than the 1000~pkd
level.
The effects due to cosmic-induced long-lived
isotopes at this shallow depth but within elaborate
shieldings
are typically at the range of a few~pkd.
The residual background can be identified,
measured and subtracted off
by various means like alpha peaks, gamma peaks,
and neutron-capture bursts.
Such background subtraction strategies
have been successfully used in accelerator
neutrino experiments.
For instance, the CHARM-II experiment
measured about 2000 neutrino-electron
scattering events from a
sample of candidate events
with a factor of 20 larger
in size~\cite{charm2},
achieving a  few \%
uncertainty in the signal rate.
A suppression factor of 100 is therefore
a realistic goal.

In addition, one can use the conventional Reactor ON/OFF
subtraction to further enhance the sensitivities.
Based on considerations above, 
a residual Background-to-Signal ratio 
of less than 10 before Reactor ON$-$OFF is attainable.
In comparison, the best published limit on
neutrino magnetic moment search with reactor neutrinos~\cite{rovno}
is based on a Si(Li) target with 
a mass of 37.5~kg at a
threshold of 600~keV and a Reactor OFF/(ON-OFF)
ratio of 120. 
Therefore, the CsI experiment should be able to achieve
a better sensitivity in the studies of
neutrino-electron scatterings.

The other applications typically allow the operation
in underground sites so that the cosmic-induced background
would be reduced compared to discussions above. 
The new challenges and complications 
are due to lower energy or smaller signal
rates, discussed as follows:
\begin{enumerate}
\item {\bf Dark Matter Searches:} \\
The present experimental background level~\cite{dmnai,dama} for the nuclear
recoil energy range (10~keV and less) is around
O(1)~pkd per keV. This is comparable to the internal radio-purity
limits already achieved in the CsI(Tl) prototype. 
However, the low energy (and therefore low light output)
makes the 
the definition of a three-dimensional fiducial volume less
efficient. 
The geometry and 
performance parameters of Figure~\ref{detdesign}
is optimized for higher energy. Nevertheless, a simple
variant of the concept can be adopted by using an
active light guide based on crystals with distinguishably
different time-profiles as the target crystals.
A possibility is the combination of pure CsI (decay time 10~ns)
and CsI(Tl) (decay time 1000~ns). 
The location of the events can be
obtained by pulse shape analysis. 
The rejection of PMT noise can be done also
by PSD but will become  delicate at the
low energy (few-photoelectrons) regime.
The background subtraction procedures 
will require detailed knowledge of the
effects from X-rays and Auger electrons
at these low energies.
Geometry of 
the crystal modules and the electronics design
should be optimized to lower the detection and PSD threshold 
as far as possible. External shieldings should
be optimized to minimize the effects of high energy neutrons 
which can penetrate easily through the active veto to give
the nuclear recoil background. 
Experiences from the operational NaI(Tl) detectors~\cite{dmnai,dama}
can provide valuable input.

\item {\bf Solar Neutrino Experiments:} \\
The energy range of interest ($>$100~keV)
allows good detector performance for crystal scintillators.
However, a much smaller $\snu$N-CC signal rate 
on the range of O(1)~per 10 tons
per day is expected.
A target mass of 
tens of tons will be required, such
that the scale-up schemes should be studied.
A major R\&D program, similar to the efforts
with liquid scintillators~\cite{ctf} to enhance - and measure -
the radio-purity level to the 10$^{-16}$~g/g range for U/Th
is necessary, for whichever target isotopes and whichever
detector techniques. 
Still, the efforts can be focussed on to a single
material, namely the crystal target itself. 
The radio-purity  requirements can be relaxed
for target isotopes which can lead to distinct
spatial and temporal signatures, like $^{115}$In~\cite{snuin}
as well as $^{176}$Yb, $^{160}$Gd and $^{82}$Se~\cite{lensgso}. 
\end{enumerate}

\section{Outlook}

Large water Cerenkov and liquid scintillator detectors 
have been successfully used in 
neutrino and astro-particle physics experiments.
New detector technology must be explored to
open new windows of opportunities. 
Crystal scintillators may be well-suited
to be adopted for low background experiments
at the keV-MeV range. 
Pioneering efforts have already been  made
with NaI(Tl) crystals for Dark Matter searches,
while another experiment with CsI(Tl) is being 
constructed to study low energy neutrino interactions.
The present O(100~kg) target mass range can be 
scaled up to tens of tons, based on the 
successful experience of calorimeters
in high energy physics experiments.

A generic detector
design is considered in this article, demonstrating
that defining a three-dimensional fiducial volume
with minimal passive materials is possible.  
The large $\gamma$-attenuation at low energy
can lead to a large suppression of 
background due to ambient radioactivity
by the active veto layers.
Consequently, the principal
experimental challenges become ones focussed on the
understanding, control and  suppression of
the radioactive contaminations in the crystals,
as well as on the optimization of the detector design
to realize an efficient, totally-active, 
three-dimensional fiducial
volume definition.
The high $\gamma$-detection efficiency,
good energy and spatial resolutions, low detection threshold,
PSD capabilities and clean alpha signatures 
provide powerful diagnostic tools towards
these ends.

There are still much room for research  and
development towards the realization of big experiments.
Potential spin-offs in other areas are possible
in the course of these efforts.

This work was supported by contracts
NSC 87-2112-M-001-034 and NSC 88-2112-M-001-007
from the National Science Council, Taiwan.

\pagebreak

\clearpage

\input{table1.tab}

\clearpage

\clearpage

\clearpage

\clearpage

\clearpage

\clearpage

\end{document}

%% file: table1.tab
\begin{table}
\begin{center}
\begin{tabular}{|l||c|c|c|c|c|}
\hline
& & & & & \\
Properties & CsI(Tl) & NaI(Tl) & BGO & Liquid & Plastic  \\
& & & & &  \\ \hline \hline
& & & & & \\
Density (gcm$^{-3}$) & 4.51 & 3.67 & 7.13 & 0.9 & 1.0  \\
& & & & & \\
Relative Light Yield & 0.45 & 1.00~$^{\dagger}$ & 0.15 & 0.4 & 0.35 \\
& & & & & \\
Radiation Length (cm) & 1.85 & 2.59 & 1.12 & $\sim$45 & $\sim$45 \\
& & & & & \\
dE/dx for MIP (MeVcm$^{-1}$) & 5.6 & 4.8 & 9.2 & 1.8 & 1.9 \\
& & & & & \\
Emission Peak (nm) & 565 & 410 & 480 & 425 & 425 \\
& & & & & \\
Decay Time (ns) & 1000 & 230 & 300 & 2 & 2  \\
& & & & & \\
Refractive index & 1.80 & 1.85 & 2.15 & 1.5 & 1.6  \\
& & & & & \\
Hygroscopic & slightly & yes & no & no & no  \\
& & & & &  \\ \hline
\multicolumn{6}{l}{ }\\
\multicolumn{6}{l}{$^{\dagger}$ Typical light yield for NaI(Tl) 
is about 40000 photons per MeV.}
\end{tabular}
\end{center}
\caption{Characteristic properties of the common
crystal scintillators and their comparison with
typical liquid and plastic scintillators.}
\label{scintab}
\end{table}